# CO-QLink: Cryogenic Optical Link for Scalable Quantum Computing Systems and High-Performance Cryogenic Computing Systems

Zheng Chang[1], Siqi Zhang[1], Wenqiang Huang[1], Tian Tian[1], Qichun Liu[3], Tiefu Li[1,3],

Nan Qi[4], Yuanjin Zheng[2], Zhihua Wang[1], Yanshu Guo[2*], Hanjun Jiang[1*]

[1]*Tsinghua University, 100084, Beijing, China*

[2]*Nanyang Technological University, 639798, Singapore, Singapore*

[3]*Beijing Academy of Quantum Information Sciences, 100193, Beijing, China*

[4]*Chinese Academy of Sciences, 100083, Beijing, China*

## Abstract

Cryogenic systems necessitate extensive data transmission between room-temperature and cryogenic environments, as well as within the cryogenic temperature domain. High-speed, low-power data transmission is pivotal to enabling the deployment of larger-scale cryogenic systems, including the scalable quantum computing systems and the high-performance cryogenic computing systems fully immersed in liquid nitrogen. In contrast to wireline and microwave links, optical communication links are emerging as a solution characterized by high data rates, high energy efficiency, low signal attenuation, absence of thermal conduction, and superior scalability. In this work, a 4K heat-insulated high-speed (56Gbps) low-power (1.6pJ/b) transceiver (TRX) that achieves a complete link between 4K systems and room temperature (RT) equipment is presented. Copackaged with a PIN photodiode (PD), the RX uses an inverter-based analog front-end and an analog half-rate clock data recovery loop. Connecting to a Mach-Zehnder modulator (MZM), the TX contains a voltage-mode driver with current-mode injection for low-power output-swing-boosting and 3-tap feed-forward equalization (FFE). This link has been demonstrated in the control and readout of a complete superconducting quantum computing system.

## 1. Introduction

### 1.1 Large-scale quantum computing systems

A fault-tolerant quantum computing (QC) system requires millions of qubits operating at cryogenic temperatures, typically in the range of hundreds or tens of millikelvin. However, the cooling power of a dilution refrigerator (DR) is very limited, making it challenging to integrate more than 100 qubits within a single DR [1]. A promising development path for large-scale cryogenic quantum computing systems lies in the adoption of multi-DR cluster architectures to scale up qubit numbers, while ensuring tight coupling between quantum processors and high-performance computing (HPC) resources. In such an architecture, whether the quantum processing unit (QPU) acts as the core—with classical HPC systems performing real-time processing and interacting with the quantum control system—or serves as a black-box accelerator within classical high-speed computations, substantial and low-latency data throughput between the quantum control system and the classical computing system is essential [2]. Furthermore, cryogenic systems across different DRs must maintain qubit coupling and correlation, as well as data interconnection between control and computing units. This gives rise to a significant demand for high-speed data links both within cryogenic systems and between cryogenic and room-temperature environments.

Specifically, cross-domain qubit coupling, storage, and transmission can be achieved using photons via optical fiber as the medium [3], [4]. For data exchange among multiple subsystems, reliable, efficient, and low-power data interfaces capable of operating under both room-temperature and cryogenic conditions, along with suitable transmission media, are required.

**1.2 High-performance cryogenic computing systems**

Currently, envisioned quantum computers are all tightly integrated and operate collaboratively with classical high-performance supercomputers. Against the backdrop of the gradual decline of Moore's Law, cryogenic computing has emerged as a potential pathway to overcome computational performance bottlenecks. Traditional room-temperature supercomputers may face challenges such as thermal management, power consumption, and the memory wall. In contrast, supercomputers operating at cryogenic temperatures could become a new solution for high-performance computing [5], [6]. This involves the design and application of various modules such as CPUs, caches, and RAM at low temperatures. Just as at room temperature, these modules—as well as different computing nodes—require high-speed data transmission when data interconnection is one of the major bottlenecks limiting computing power. Although the power consumption of links does not account for the main portion compared to that of computing units, high-energy-efficiency links can still significantly reduce costs and increase computational power density. This necessitates the implementation of data links and interfaces that can operate stably under cryogenic conditions. Such data links are required for future high-performance cryogenic computing systems that are fully immersed in 77K liquid nitrogen [5], [6], [7].

**1.3 Methods of cryogenic data link**

As shown in [8], a 300K–4K/4K–300K link achieving 40Gb/s has been demonstrated in a cryogenic probe station. However, the use of a single SC-086/50-SCN-CN coaxial cable introduces approximately 0.35mW of passive heat load to the 4K stage and 7mW to the 50K stage in a Bluefors XLD1000-SL DR [9]. Although reducing the diameter of coaxial cables has proven effective in mitigating heat conduction—decreasing passive heat load with the square of the diameter—this approach comes with a critical trade-off: the attenuation and loss of the coaxial cable increase linearly [1]. While heat conduction is reduced, achieving the same data rate necessitates either a greater number of coaxial cables or increased power consumption at the cryogenic transceiver interfaces for signal recovery and equalization. This indirectly amplifies the cooling burden on the refrigerator, consuming a significant portion of its power budget.

Additionally, a THz microwave backscattering data link has been proposed for data transfer between room temperature (RT) and 4K, but its data rate is limited to below 10Gb/s [10]. Moreover, the THz wireless link relies on a physical aperture at the top of the DR, resulting in poor scalability and limited transmission distance. For more complex, large-scale quantum computing systems, THz links are inadequate to meet the demands for sophisticated, high-throughput, and multi-directional data transmission.

In contrast, optical links are preferred due to their high bandwidth, minimal channel loss, excellent scalability, and strong thermal isolation. Several photonic devices have been tested at 4K or even lower temperatures to validate the feasibility of optical links [11], [12], [13], [14], [15], [16], [17], [18], [19], [20], but no fully integrated high-speed optical transceiver has been demonstrated, nor have results been reported from co-testing with quantum computing systems.

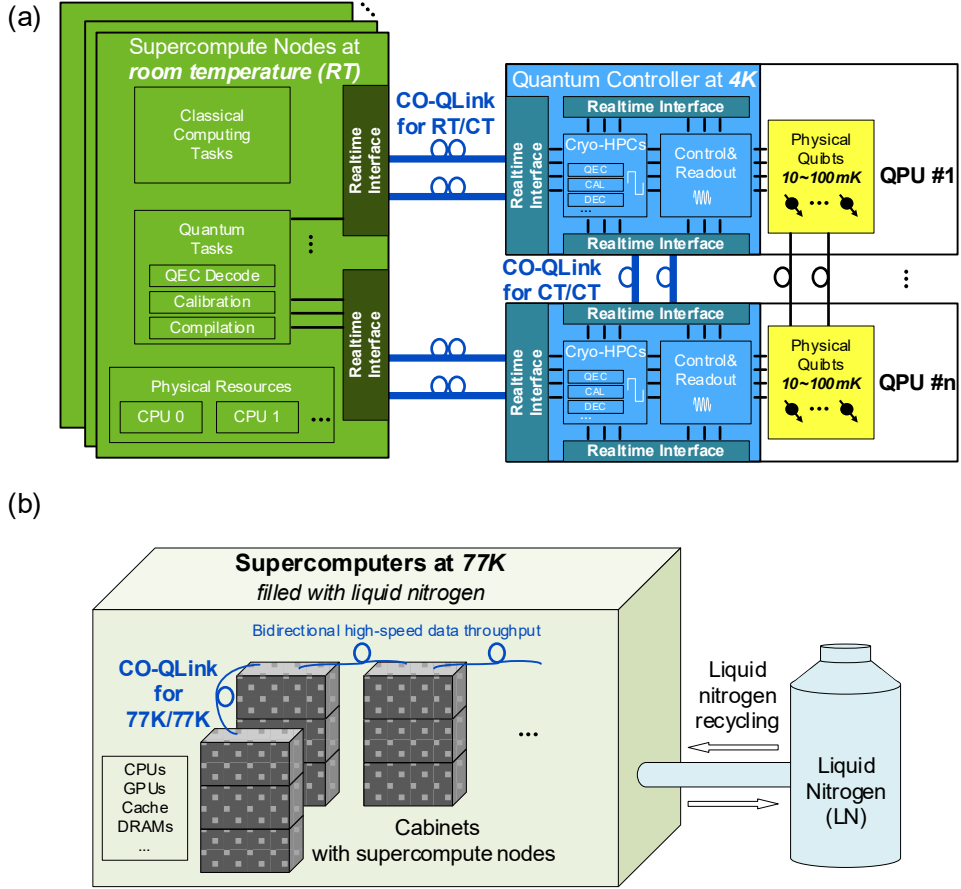

Fig. 1. (a) CO-QLink in the supercomputer with quantum processors; (b) CO-QLink in the high-performance cryogenic computing system.

This work presents a full up-to-56Gb/s cryogenic optical link (CO-QLink), targeting scalable high-performance quantum systems with tightly-integrated classical supercomputers and QPUs, as shown in Fig. 1(a), and high-performance cryogenic computing systems fully immersed in liquid nitrogen, as shown in Fig. 1(b). This link is capable of RT/CT (Cryogenic temperature) and CT/CT data transmission. As shown in Fig. 2(a), an opto-electrical RX ASIC and an opto-electrical TX ASIC operating at the 4K stage have been designed and implemented within this data link to facilitate the reception and transmission of data at cryogenic temperatures. Large-scale cryogenic systems incorporating multiple DRs require the establishment of the transmission and sharing of data, such as quantum instructions, as well as the coupling between qubits for QC applications, across different cooling units. These requirements can be effectively addressed through fiber-based approaches. The present work primarily focuses on achieving high-speed and energy-efficient data transmission between such systems.

## 2. Link implementation

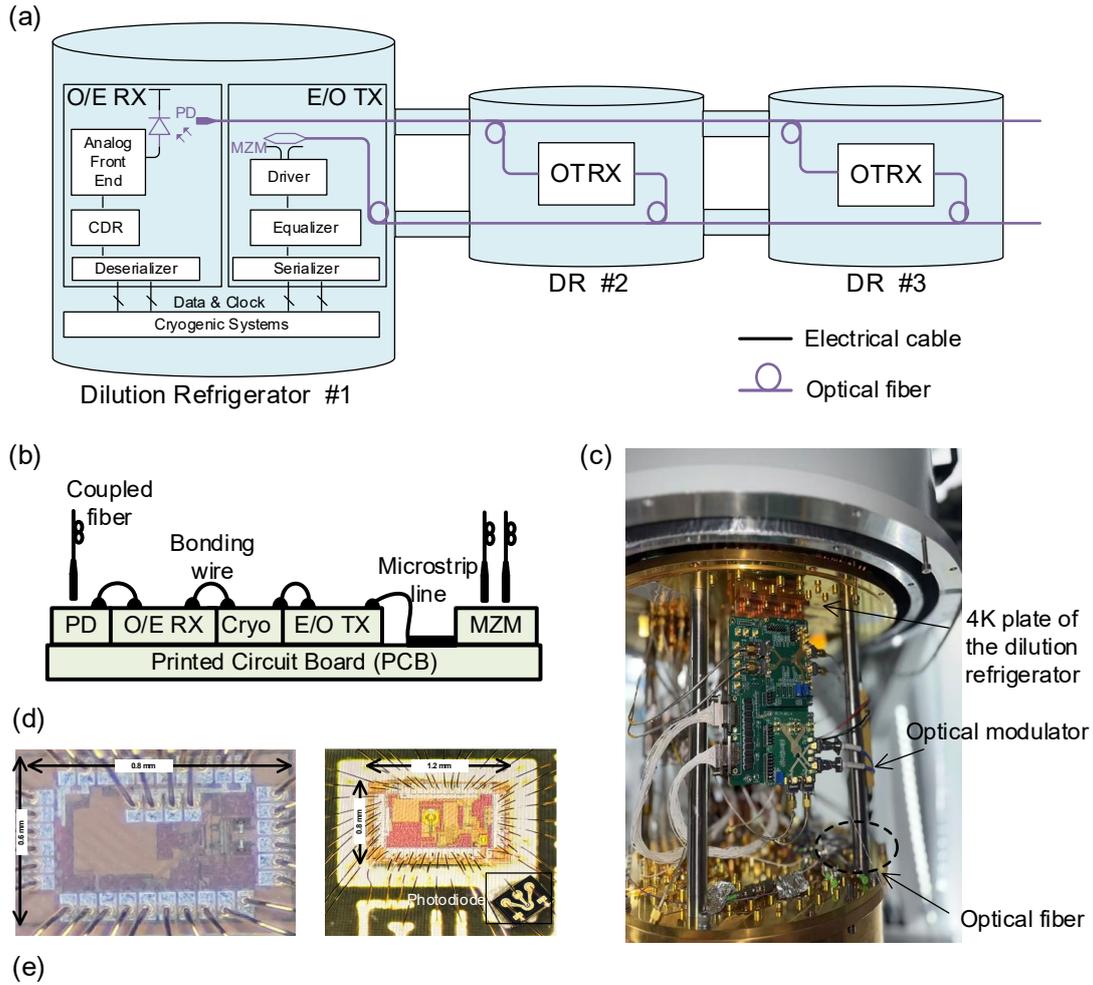
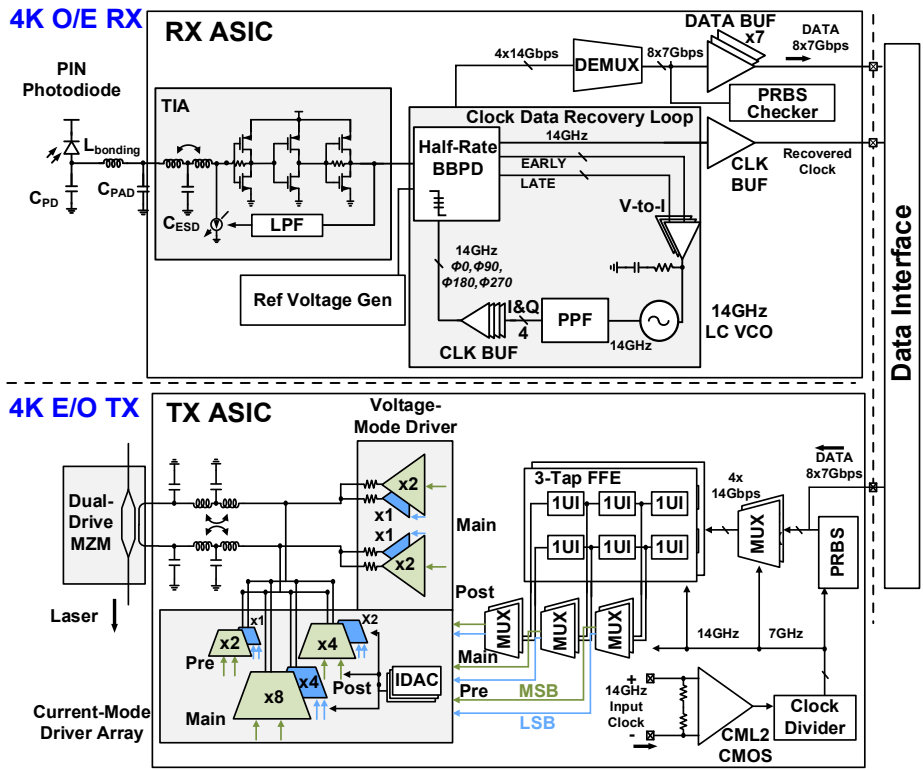

Fig. 2. (a) Proposed optical link with cryogenic opto-electrical transceiver (OTRX); (b) Proposed cryogenic OTRX implementation; (c) OTRX mounted at the 4K plate of a Bluefors DR; (d) Die photos of the OTRX fabricated in 28nm CMOS; (e) Cryogenic opto-electrical link with RX ASIC (top) and TX ASIC (bottom).

The architecture of the optical link with cryogenic OTRX is shown in Fig. 2. A room-temperature laser source drives the 4K–4K cryogenic direct optical link using a 1530nm C-band laser. Each DR in the large-scale QC system is interconnected via optical fiber, enabling broadband data transmission without thermal conduction. Inside each DR, a pair of E/O TX and RX modules is integrated. The O/E RX, co-packaged with an InGaAs PIN photodiode (PD), demodulates up to 56Gb/s data rate with PAM-4 modulation and deserializes it into eight parallel outputs for direct use by DACs in qubit control/readout ASICs or for transmission through an E/O TX over longer distances. The E/O TX integrated a LiNbO$_3$ Mach–Zehnder modulator (MZM), serializes parallel data from readout ADCs, and modulates them onto the C-band laser.

Cryogenic characteristics of commercial LiNbO$_3$ MZMs and standard InGaAs/InP PIN PD have been studied in existing works [11], [14], [19]. Unlike conventional VCSELs, which suffer from performance mismatch and operational failure at cryogenic temperatures, or thermally sensitive devices such as micro-ring modulators, conventional Mach-Zehnder modulators (MZM) and photodiodes (PD) remain functional under low-temperature conditions while retaining certain performance characteristics. The half-wave voltage of MZMs undergoes variation under cryogenic conditions, and photodiodes exhibit a blue shift in their response spectrum due to changes in the bandgap[14], [19]. Utilizing components that demonstrate stable performance at both room temperature and cryogenic temperatures can effectively enhance system robustness and reusability, while simultaneously reducing costs.

Schematics and die photos in Fig. 2(c), Fig. 2(d), and Fig. 2(e) show the detailed topologies of the proposed optical link TRX. The O/E RX consists of an InGaAs PD and an RX ASIC. The RX ASIC comprises a transimpedance amplifier (TIA), a clock data recovery loop (CDR), and a de-serializer. The PD has a responsivity of 0.35A/W and capacitance of 80fF at 4K. The PD accepts a C-band laser (1530nm) and generates a reverse current fed into the TIA through a 1mil bonding wire. A T-coil is used to compensate for the parasitic capacitance of ESD circuits. To adapt to the increasing threshold voltage at 4K and fully utilize the gm of MOSFETs, a three-stage inverter-based Cherry-Hooper TIA is implemented. The TIA output is connected to an inverter-based low-pass filter (LPF), and the LPF controls a variable current source to realize DC compensation. The analog front-end has a transimpedance of 2.1kΩ, and its output is fed to a half-rate bang-bang phase detector (BBPD), which is triggered by four quadrature 14GHz clocks. The phase detector generates EARLY and LATE signals to drive the voltage-to-current (V–I) converters, which, after passing through a loop filter, control the varactors of a 14GHz LC voltage-controlled oscillator (VCO). The VCO employs a 5-bit capacitance bank for tuning. The measured tuning range is from 13.5GHz to 14.6GHz. This CDR loop directs the sampling clocks to the optimum positions. And the bang-bang phase detector samples 4x 14Gb/s data by quadrature clocks, and after an array of 2:1 DEMUX cells, the 56Gb/s (8x 7Gb/s) data stream can be recovered. The 8x 7Gb/s data are also fed into a pseudo-random bit sequence (PRBS) checker for bit error rate (BER) measurement.

The E/O TX includes a TX ASIC and an MZM. The TX ASIC supports a data rate of 56Gb/s (8x 7Gb/s) with an on-chip PRBS generator used for chip performance measurement. The 8x 7Gb/s data are then serialized by a 2:1 MUX array. In order to compensate for the channel loss of the modulator and the transmission line connecting the modulator to the electrical chip, the output 4x 14Gb/s data are then fed into a 3-tap generation circuit based on 1-UI delay latches for feed-forward equalization. The PRE, MAIN, and POST sequences are then serialized by a half-rate 2:1 MUX array. Considering the advantage of the Vπ of long lithium niobate modulators and the need for low power consumption, the TX employs a monolithic dual-drive LiNbO3 Mach-Zehnder modulator with a low Vπ and high extinction ratio at the cost of volume. The dual-drive mode not only saves power consumption at a certain driving voltage but also cancels common-mode noise from the chips and the environment. To generate a sufficiently large driving voltage amplitude and save as much power consumption as possible simultaneously at the 28nm node, a source-series-terminated (SST) PAM-4 driver array, which typically consumes ~4x lower power compared with the current-mode driver for the same output swing, along with a PMOS current-mode-logic (CML) driver array broadening the limit of the supply voltage is employed. The CML array also realizes the 3-tap FFE by adjusting the current bias of PRE, MAIN, and POST driver cells using IDACs. The differential output swing reaches 1.3V.

## 3. Measurement results

### 3.1 Chip performance at 300K and 4K

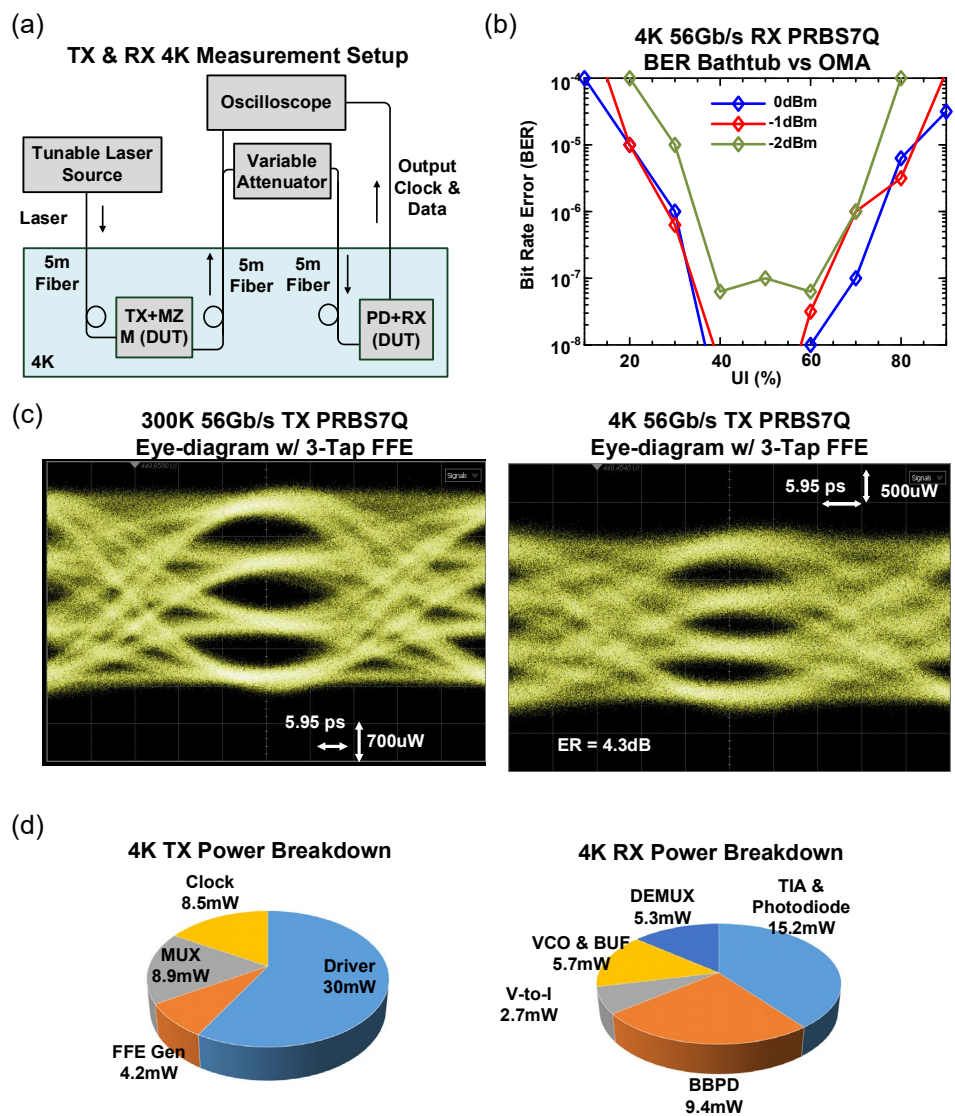

Fig. 3. (a) Opto-electrical link measurement setup; (b) The BER bathtub results of 56Gb/s RX at 4K with different OMA; (c) The 56Gb/s optical output eye-diagram of TX at 300K and 4K with PRBS7Q and 3-TAP FFE; (d) The power breakdown of OTRX working at 4K.

The two ASICs were fabricated in 28nm bulk CMOS technology. The O/E link measurement setup is shown in Fig. 3(a). The under-test O/E RX and E/O TX are laid at the 4K plate of a dilution refrigerator. A 1530nm laser generated by a C-band laser source at RT is led to the input port of the MZM modulator through a 5m fiber. The fiber crosses temperature zones through the FC/APC interfaces at the top of the refrigerator. Fig. 3(c) shows the 56Gb/s E/O TX optical output eye diagram at 4K with an on-chip PRBS7Q generator, achieving an optical output extinction ratio of 4.3dB. After the E/O TX measurement, the optical output of E/O TX is connected to the O/E RX photodiode by a variable optical attenuator. With the on-chip PRBS7Q generator and checker, both links are capable of <1E-8 bit error rate (BER). By disabling the CDR loop and switching the sampling clock to be generated by an off-chip 14GHz signal source, the 4K O/E RX's BER bathtub is shown in Fig. 3(b). The O/E RX working at 4K meets <1E-8 BER output for >18% UI opening with >-1dBm optical modulation amplitude (OMA) input at 56Gb/s.

**3.2 Demonstration in superconducting quantum computing systems**

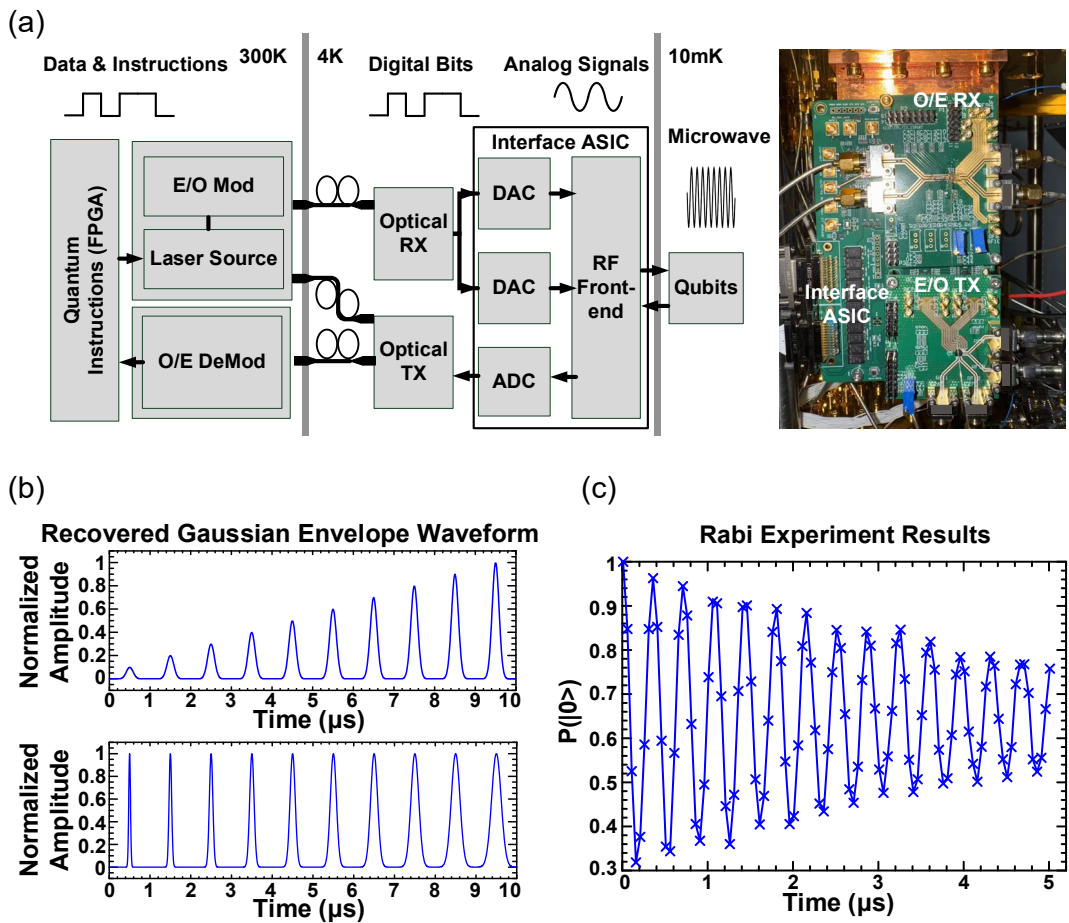

Fig. 4. (a) Quantum computing system prototype with proposed optical link and its measurement setup; (b) Gaussian envelope waveform recovered by the O/E RX for qubit control; (c) Rabi experiment results.

To further verify the 4K O/E link, the measurement setup shown at the top of Fig. 4(a) was built. An E/O TX evaluation board and an O/E RX board are connected to a cryogenic qubit interface ASIC. The O/E RX recovers the data stream from the optical signal and, through the DAC in the interface ASIC, generates Gaussian-envelope microwave driving pulses with varying amplitude or duration. The E/O TX receives data from the interface ASIC and modulates the optical signal for long-distance transmission. A data interface block integrated in the ASIC acts as a FIFO, enabling data transfer between the 8-bit OTRX and the 12-bit DAC or 9-bit ADC in the qubit interface ASIC. The recovered Gaussian-envelope waveform is shown in Fig. 4(b). Under the control of the optical link and the interface ASIC, a Rabi oscillation experiment is successfully demonstrated (Fig. 4(c)).

## 4. Conclusion

| Reference | | This Work | JSSC'24[8] | Nat. Electron.'25[10] |
|---|---|---|---|---|
| Methods | | Optical | Wireline | THz |
| ASIC Technology | | 28nm CMOS | 40nm CMOS | 22nm FinFET CMOS |
| Modulation | | PAM4 | PAM4 | OOK/BPSK |
| Data Rate @4K | | 56Gb/s | 40Gb/s | 4Gb/s(TX), 4.4Gb/s(RX) |
| TX Power @4K | Front-End | 30mW, 0.54pJ/b | 28.7mW, 0.72pJ/b | 0.86mW, 0.21pJ/b |
| | Data & Clock Path | 21.6mW, 0.39pJ/b (4:1 MUX) | 69.9mW, 1.75pJ/b (64:1 MUX) | / |
| RX Power @4K | Front-End | 15.2mW, 0.27pJ/b | / | 0.15mW, 0.03pJ/b |
| | Data & Clock Path | 23.1mW, 0.41pJ/b | / | / |
| Heat Transfer | | Extra Low | High | Extra Low |
| Link Reach | | Long (>5m) | Medium (~1m) | Short (~10cm) |

Tab. 1. Link performance summary and comparison with other existing methods

Tab. 1 summarizes the performance of this link and compares it with state-of-the-art methods. This work demonstrates the first 4K–4K direct OTRX link for scalable quantum computing, enabling cryogenic communication crossing multiple dilution refrigerators. This is the first cryogenic optical full data link that can be directly integrated into the cryogenic systems and has been demonstrated in a practical quantum computing system. The link achieves 56Gb/s with energy efficiencies of 0.92pJ/b (TX) and 0.68pJ/b (RX) at 4K. Compared to THz microwave and wireline links, it promises to support potentially a much longer reach and offers potential for higher rates and better efficiency with WDM and Si-photonics.